\documentstyle[preprint,eqsecnum,aps,floats]{revtex}

\begin{document}
\draft
\title{   The Transition between Immune and Disease States in a Cellular
Automaton Model of Clonal Immune Response}
 
\author{Michele Bezzi}
\address{Dipartimento di Fisica, Universit\`a di Bologna, Via Irnerio
46, 40126 Bologna, Italy}
\author{Franco Celada}
\address{Hospital for Joint Diseases, 301 E. 17th.  St., New York, NY 10598,
USA and Cattedra di Immunologia, Universit\`a di Genova, Largo
Rosanna Benzi 10, 16132 Genova, Italy}
\author{Stefano Ruffo}
\address{Dipartimento di Energetica "S.Stecco", Universit\`a di
Firenze, Via S.Marta 3, 50139 Firenze, Italy.   INFN and INFM, Firenze, Italy}
\author{Philip E. Seiden}
\address{IBM Thomas J. Watson Research Center, P.O. Box 218,
Yorktown Heights, New York 10598, USA}
\date{\today}
\maketitle
\begin{abstract}
In this paper we extend the Celada-Seiden (CS) model of the humoral
immune response to include infectious virus and cytotoxic T lymphocytes
(cellular response). The response of the system to virus involves a
competition between the ability of the virus to kill the host cells and
the host's ability to eliminate the virus.  We find two basins of
attraction in the dynamics of this system, one is identified with disease
and the other with the immune state.  There is also an oscillating
state that exists on the border of these two stable states.
Fluctuations in the population of virus or antibody can end the
oscillation and drive the system into one of the stable states.  The
introduction of mechanisms of cross-regulation between the two
responses can bias the system towards one of them.  We also study a
mean field model, based on  coupled maps, to investigate virus-like
infections.  This simple model reproduces the attractors for
average populations observed in the cellular automaton.
All the dynamical behavior connected to spatial extension is lost, as
is the oscillating feature.  Thus the mean field approximation
introduced with coupled maps destroys oscillations.
\end{abstract}
 
\narrowtext
 
\section{Introduction}
 
The immune system is the body's defense against attacks from foreign
substances (called {\it antigens\/}) such as parasites, bacteria and
viruses.  It has two different branches to fight against specific
antigens: the {\it humoral response\/} (mediated by antibodies) and
the {\it cellular\/} response (cell-mediated).  In this paper we
present a cellular automaton model of how these two parts of the
immune system interact with each other during the response to antigenic
stimulation.
 
Celada and Seiden \cite{celada1,celada2} have proposed a model for
the humoral response based on cellular automata.  The aim of this
model is not to simulate the immune response in all its features,
which is clearly an impossible task.  Although it is impossible to
replace {\it in vivo\/} or {\it in vitro\/} experiments with computer
simulations, computer simulations may be a useful tool in performing
{\it in machina\/} experiments, which can then be used to design the
more expensive and time consuming {\it in vivo\/} and {\it in vitro\/}
experiments.
 
The organization of the paper is the follows: in Sec.\ \ref{sec:biol}
we give a short account of the biological background, for a more
complete description of the immune system see \cite{abbas}.  In Sec.\
\ref{sec:model} we sketch the main features of the Celada-Seiden model
for humoral response.  In Sec.\ \ref{infect} we discuss our extension
of the CS model which introduces cellular response into the automaton
model, and we present the results of our computer simulations.  In
Sec.\ \ref{sec:maps} a system of {\it coupled maps\/} is proposed to
describe the time evolution of populations in the immune system during
an infection.  In this section we show the results of analytical and
numerical analysis of this model and discuss the similarities and the
differences between this approach and the cellular automata.
 
\section{Biological features}
\label{sec:biol}
 
The defense mechanisms used by the body against an attack from
antigens are several, they include: physical barriers, phagocytic
cells, different clones of particular white blood cells ({\it
lymphocytes\/}) and various blood-borne molecules (e.g., {\it
antibodies\/}).  Some of these mechanisms are present prior to
exposure to antigens and their action is non-specific,
i.e., they do not discriminate between different antigens and their
response doesn't change upon further exposure to the same antigen.
This kind of response is called {\it natural immunity\/}.  There are
other mechanisms with more specific behavior, they are induced by
antigens and their response increases in magnitude and defense
capabilities with successive exposures to the same antigens.  These
mechanisms are called {\it acquired\/} (or {\it specific\/}) {\it
immunity\/} and are what we consider in this paper.
 
The main features of specific immunity are:
 
\begin{itemize}
\item {\it Specificity\/}: the immune response is highly specific for
distinct antigens, that is only a small number of particular membrane
receptors recognize a given antigen.  These receptors
bind specific portions of the antigen (called {\it antigenic
determinants\/} or {\it epitopes\/}).  Only a small fraction of the
system recognizes any particular epitope, and cells which do recognize
antigenic epitopes are positively selected by this recognition process
so that their population increases ({\it clonal selection\/}).
 
\item {\it Diversity\/}: the possible number of lymphocytes with
different specificity (different {\it clones\/}) is extremely large
($\geq 10^9$), although the total diversity actually existing in an
animal is much smaller.  The diversity is due to the variability of
lymphocyte receptors, which derives from the mechanism of genetic
recombination that occurs during cell development.
 
\item {\it Maturation\/}: the response evolves by increasing the
average affinity to the antigen through competition and selection for
binding following mutation of the genes coding for the B-cell receptor.
 
\item {\it Memory\/}: exposure of the immune system to an antigen
enhances, in quality and quantity, its capability to respond to a
second exposure to the same antigen ({\it secondary response\/}).
This is due mainly to the persistence, after any response, of
re-stimulable cells (memory cells) ready to mount a new response to
the same antigen.
 
\item  {\it Discrimination of self from non-self\/}: the immune system
doesn't respond to substances produced by the human body ({\it tolerance
to self\/}).
\end{itemize}
 
The specificity of the immune response is due to a class of white
blood cells called lymphocytes.  These are not able to begin a
response without the help of various types of cells known as {\it
antigen presenting cells\/} (APC), such as macrophages, dendritic
cells, etc. Lymphocytes are present in the blood, lymph and lymph
nodes; in the human body there are about $10^{10}$ lymphocytes.  The
two major classes are: {\it B\/} and {\it T lymphocytes\/}.
 
In mammals B lymphocytes (so called because in birds they are produced
in an organ called the {\it Bursa Fabricii\/}) mature in the bone
marrow, they are able to bind antigen and to produce antibodies with
the same specificity as their membrane receptors.  Upon binding
antigen a B cell will endocytose the antigen into small pieces between
about 8 to 15 amino acids long (peptides).  These pieces are bound to
surface receptors called major histocompatibility molecules (MHC class
II).  Helper T cells are able to recognize and bind to these
MHC/peptide complexes.  When they do they trigger T-and B-cell
proliferation and differentiation into memory cells and {\it plasma
cells\/} (which produce antibodies).
 
T lymphocytes are born in the bone marrow, and then migrate to the thymus
where they mature and are selected. Here a lot of T lymphocytes are
destroyed to avoid auto-immune responses ({\it clonal deletion\/}). There
are two different types of T lymphocytes which are important for our
model:
 
\begin{itemize}
\item {\it helper T cells\/} (Th) which are involved in B-cell
activation; they have receptors that can bind MHC/peptide
complexes.  As a consequence of this process, they produce
a set of particular molecules, called {\it cytokines\/}(e.g.,
interleukin 2 and 4), that activate B and T cells;
 
\item {\it cytotoxic T lymphocytes\/} (CTL) which can
recognize, in an antigen specific way, cells infected by viruses, and
kill them by lysis.  The recognition is similar to that of helper T
cells in that pieces of antigen are presented on another type of
MHC molecule (class~I).
 
\end{itemize}
 
Specific immune responses are classified into two different types,
based on the components of immune system that mediate the response.
 
\begin{itemize}
\item {\it Humoral response\/}, where B cells recognize antigens and
produce antibodies that attack them.  All these processes are
mediated by Th cells.  This is the main defense against
extra-cellular microbes.
 
\item {\it Cellular response\/}, mediated by CTLs which recognize
infected cells and eliminate them (help is also needed from Th
cells).  This response is effective against intra-cellular viruses.
\end{itemize}

\section{A model for the humoral response.}
\label{sec:model}
 
A cellular automaton model implementing the humoral response (to be
called CS herein) has been recently proposed \cite{celada1}.  A
cellular automaton is a dynamical system with the following
characteristics \cite{wolfram}:
 
\begin{itemize}
\item  it has a discrete number of sites;
 
\item  it evolves in discrete time steps;
 
\item  each site can take a finite number of values;
 
\item  the evolution rule is deterministic;
 
\item the evolution rule of a site depends only on a local
neighborhood of sites around it.
\end{itemize}
 
\noindent
The CS automaton modifies and extends these rules, allowing probabilistic
evolution, making the evolution rules depend only on entities on the same
site and permitting entities to move to neighboring sites. These are
typical features of reactive lattice gases (see \cite{Kapral96} for a
review). The CS automaton is defined as a two
dimensional lattice, usually of a small size ($15\times 15)$, that
represents a small part of the body.
 
\subsection{Components of the model system}
 
An important characteristic of the CS model is the simulation of
diversity in antigens and response by the introduction of several
clonotypic elements (e.g., epitopes, peptides, and receptors)
represented by binary strings.  The objects present in the model are:
 
\begin{itemize}
\item  cells (APC, B cells and T cells);
 
\item  antibodies;
 
\item  antigens.
\end{itemize}
 
B and T cells are represented by different clones, each clone is
characterized by its surface receptor which is modeled by a binary
string of $N$ bits with a fixed directional reading frame.  In
Fig.~\ref{cells}a we show how these objects are represented in the
model.  Each clonotypic set of cells has a diversity of $2^N$; for the
simulations described in this paper we use $N=8$, so we have a
diversity of $256$ (in comparison to $\sim 10^9$ of the real system).
The number of possible states in a site is therefore very high, for
example if the maximum number of cells of each kind is $1000$, the
number of states in each site is $\sim 10^{1536}$.
 
Besides the receptor (BCR), B cells have MHC class II molecules on
their surface.  In the model they are also represented by a binary
string of $N$ bits.  MHC's are involved in the process of B-cell
activation.  MHC diversity in a given body is very small, less than
$10$ different MHC molecules are present.  We have generally used one
or two different kinds of MHC molecules in the simulations.
 
We have used a complete repertoire for B cells, i.e., all can be
produced by the system and, for the parameters we use here, the
average occurrence of each clonotype in the starting population is
approximately one.  For T cells we begin with the complete repertoire,
but they are filtered by the {\it thymus\/} before putting them in the
lattice.  The thymus is an organ through which T cells must pass
before they can mature.  In the thymus they are exposed to
self-antigens presented on APCs.  If they respond too strongly or not
at all they are killed.  Therefore, the thymus acts as a filter to
remove dangerous or non-responsive T cells.  It is the body's first
line of defense against autoimmune disease.  Including a thymus,
results in a T-cell population that has restricted diversity.
 
APCs represent non-specific antigen presenting cells (such as
macrophages), they have no specific receptor (they can bind any type
of antigen with a fixed probability) but they have the same MHC as B
cells.
 
Antigens (Ag) are made of two different parts: {\it epitopes\/} and
{\it peptides\/}.  The epitope is the portion of an antigen that can
be bound by the BCR; after this event the peptide (or better a peptide of
the antigen, because generally an antigen has more than one peptide or
epitope) is presented on an MHC molecule for T-cell cognate
recognition (we show a sketch of the process in Fig.~\ref{cells}b).
 
Antibodies (Ab) are also made of two parts.  They have a receptor
({\it paratope\/}), that is represented by the same string as the BCR
of the parent B cell which produced them and, optionally, a peptide
({\it idiopeptide\/}).
 
\subsection{Interactions }
 
There are precise interaction rules; the allowed interactions are
mainly of two types:
 
\begin{itemize}
\item[-] {\it specific\/}, as between antigen and antibody, antigen
and BCR, or MHC-peptide complex and T cell receptor; this kind of
interaction has a probability of interaction (affinity) evaluated
according to the number of complementary bits between the binary
strings that represent the receptors.  In our simulations we have used
a probability of $1$ in case of perfect complementary, $0.05$ for the
case of one mismatching bit, and $0$ for more than one mismatching
bit.
 
\item[-] {\it non-specific\/} as APC-antigen interaction.  The
interaction takes place with a fixed clone-independent probability,
typically equal to $0.002.$
\end{itemize}
 
The main process of humoral response is B- and T-cell
proliferation (clonal
growth) and consequent antibody production.  The process is divided
into four parts: antigen-BCR interaction, endocytosis and peptide
presentation on MHC, T cell recognition of the MHC/peptide complex,
and finally, cell proliferation and differentiation into plasma
cells (which produce antibodies) and memory cells (which provide the
possibility of a further enhanced response).  For each antigen several
B-cell clones have receptors able to bind it.  After
binding, the B cell processes the antigen and presents it on its MHC
and then waits for a recognition/binding event by a T cell.  This is
the signal for the B cells to proceed in the response.  The signal may
fail to be delivered either because the specific T cells have been
negatively selected in the thymus, or because the probability of
finding the right combination of B and T cells in the same site is
low.  The latter situation may change with time, e.g.,
after the T cells
have proliferated through stimulation by APC presenting the same
antigen.  After they have received this second signal B cells can divide,
producing memory and plasma cells.  The plasma cells secrete
antibodies, having the same receptor as the B cell, which can bind the
antigen just as the BCR does.  Upon T-B cell interaction the T cells
also proliferate.
 
\subsection{Simulations}.
 
During a time step each site is considered individually.  Each entity
in a site is given an opportunity to take part in all interactions for
which it is able.  The success or failure of an interaction is
determined by comparison of its probability with a random number.
Although all possible interactions are considered, an entity can have
at most one successful interaction on any one time step.  After the
interactions are determined, the entities are allowed to die (with some
half life), stimulated cells divide, new cells are born, and
antibodies are generated.  Finally, the entities are given an
opportunity to diffuse to neighboring sites.  This constitutes a time
step and the entire process is repeated for as many time steps as
desired.
 
To give a rough idea of the functioning of the CS model we show a
typical simulation of {\it immunization\/}
in Fig.~\ref{immsim}.
   The system initially has
no antigen, no antibodies, and $1000$ B cells, $1000$ T cells and
$1000$ APCs uniformly distributed in space and receptor type
(except that the T cells have been processed in the thymus); no
interaction among species is present and the system is in a steady
state where the natural death rate equals the birth rate.  We start by
injecting a single type of antigen (same epitopes
and peptides).  The antigens are at first bound primarily by the APCs
since, although weakly binding, they are much more plentiful than the
rare B cells that match the antigen.  T cells will stimulate these
APCs and then divide to form populous clones.  There will then be
sufficient antigen responding T cells to easily find the few
responding B cells.  Upon stimulation from the T cells the B cells
also divide and form significant clones.  Finally, antibodies are
produced and the antigen is removed.  This is the {\it primary
response\/}.
 
If later more of the same antigen is injected (e.g., at time
step 100 of Fig.~\ref{immsim}) it is removed much more rapidly because
the system has an appreciable population of B and T memory cells
induced by the primary antigen dose.  This is the {\it secondary
response\/} and it can be so strong and swift that the antigen is
eliminated before it can do any damage.
 
A number of other aspects of the humoral immune system have been
studied with this model, e.g., response to various levels of
antigen dose \cite{celada1}, hypermutation and affinity maturation
\cite{hyper}, and thymus function \cite {thymus}.
 
\section{Infection and CTL response.}
\label{infect}
 
We have introduced three new features into the CS model:
 
\begin{itemize}
\item  {\it infection\/}
 
\item  {\it CTL response \/}
 
\item  {\it cross regulation\/}
\end{itemize}
 
The aim is to implement some mechanisms of cross regulation and study
how the cellular and humoral responses interact with each other.  The
CTL response is activated against intracellular viruses or parasites,
so before introducing CTL we need an infection step.  Most of our
simulations are limited to understanding the features of infection.
 
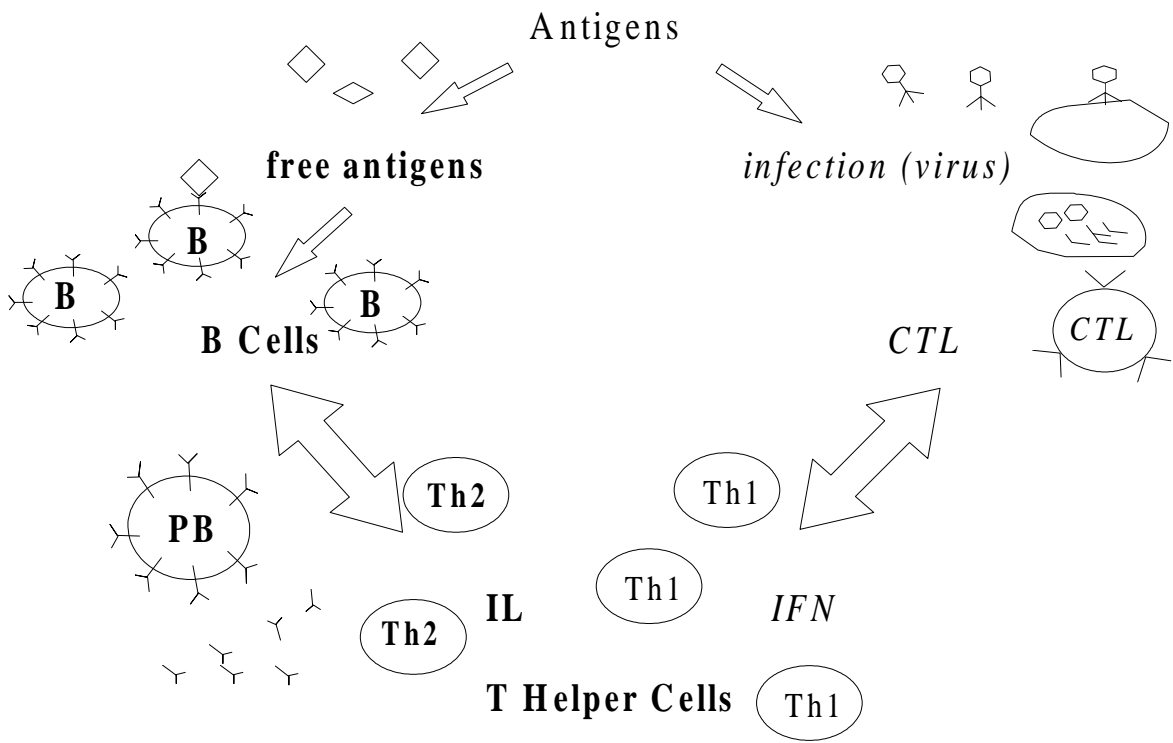
\begin{figure}
\caption{A scheme of parallel humoral and CTL response with
effectors and regulators.  Bold type indicates the part of the
response simulated by the basic Celada-Seiden model, italics the
part simulated in the extended model.}
\label{ctls}
\end{figure}

In Fig.~\ref{ctls} we sketch the processes modelled in CS model (in bold)
and the modifications we have added to implement CTL response (in italics).
 
\subsection{Infection}
 
We consider one kind of antigen, a virus with one epitope and with
infective function, i.e., capability to penetrate cells and to multiply
inside them, like a virus or an intracellular parasite.
All references to antigen from now on refer to
this virus.  It has a given probability $P_i$ per time step to
infect any B cell or APC present in the same site.  The target cells
are the cells of immune system: B cells and APC.  We chose to infect
only these to limit the cell species in the model but infections of
this type are known, for example, {\it Epstein-Barr \/}virus which
infects B cells and {\it Leishmania major\/} an intracellular
parasite of macrophages.
 
Infected cells continue their {\it normal\/} life while the virus
duplicates inside them with a constant growth rate $\left( r\right)$
(this may be a drastic approximation of realistic situations). That is the
number of viruses inside the cell at time $n$ ($V_I^n$) grows according to:
 
\[
V_I^{n+1}=r\times V_I^n
\]
 
When the number of viruses inside the cell exceeds a fixed threshold
($V_{\max }$), the cell is destroyed and $V_{\max }$ viruses are
freed.  The virus is shielded from antibodies when it is inside the
cell but can be destroyed when it is outside.  In our first
simulations, featuring only humoral response, virus inside the cells
is safe; later we will consider the case where lymphocytes (CTL)
can recognize and kill infected cells.
 
A first set of simulations was performed by introducing a single
injection of virus and then observing the response of the system as
a function of $V_{\max }$ and $P_i$.
 
We have observed three different final states:
 
\begin{itemize}
\item  indefinite growth of the virus ({\bf V}, diseased state), see
Fig.~\ref{evol} (right-hand panels);
 
\item  elimination of the virus ({\bf IS}, immune state), see
Fig.~\ref{evol} (left-hand panels);
 
\item  an oscillatory state ({\bf O}), see
Fig.~\ref{evol} (middle panels).
\end{itemize}
 
Fig.~\ref{evol} presents the total number of B cells and virus versus
time.  In Tables \ref{su2} and \ref{su1} we summarize the
results of the simulations.  There are two fixed point type basins of
attraction
{\bf IS} and {\bf V}.  The {\bf O} state appears at the border between
these two.
 
\begin{table}[tbh]
\caption{Final state for various values of  $P_i$ and $V_{\max }$}
\label{su2}
\begin{tabular}{|l|c|c|c|c|}
& $P_i=0.1$ & $P_i=0.05$ & $P_i=0.025$ & $P_i=0.01$ \\ \hline
$V_{\max }=8$ & {\bf V} & {\bf IS} & {\bf IS} & {\bf IS} \\ \hline
$V_{\max }=12$ & {\bf V} & {\bf O} & {\bf IS} & {\bf IS} \\ \hline
$V_{\max }=16$ & {\bf V} & {\bf V} & {\bf V} & {\bf IS} \\ \hline
$V_{\max }=25$ & {\bf V} & {\bf V} & {\bf V} & {\bf O} \\ \hline
$V_{\max }=32$ & {\bf V} & {\bf V} & {\bf V} & {\bf V} \\
\end{tabular}
\vspace{0.2in}
\end{table}

\begin{table}[tbh]
\caption{Final state for $P_i$$=0.05$}
\label{su1}
\begin{tabular}{|l|c|c|c|c|c|c|c|c|c|c|c|}
{$V_{\max }$} & 6 & 7 & 8 & 9 & 10 & 11 & 12 & 13 & 14 & 16 & 32 \\
\tableline
Final state & {\bf IS} & {\bf IS} & {\bf IS} & {\bf O} & {\bf O} &
{\bf O} & {\bf O} & {\bf V} & {\bf V} & {\bf V} & {\bf V } \\
\end{tabular}
\vspace{0.2in}
\end{table}

If we look at the antigen concentration in the {\bf O} case we see low
numbers of B cells correspond to high values of virus, i.e.,
they are in phase opposition . The population of virus is extinguished
when the concentrations of free and intracellular virus are so low that
they cannot infect enough cells before being eliminated by antibodies.
So at low concentration of antigen, it may happen that there is not
enough at some time to restart the process of infection; this
leads to the elimination of antigen and to the recovery of the initial
concentration of B cells ({\bf IS}).  However the opposite process is
also possible: in the high-antigen concentration phase, virus can kill
most of the clones that are able to initiate the immune response.  So
after a number of oscillations, it is possible to find that the system
suddenly relaxes to the diseased state ({\bf V}).  Fluctuations in the
number of virus in the former case, or in the number of B cells in the
latter case, can destroy the oscillating mode.
 
The period of oscillation is much larger than the characteristic time
of infection and the latency period (i.e., the time in which
virus grows inside an infected cell).  In fact, in our simulations, the
latency period is about 10 time steps while the oscillation period is, at
least, ten times larger.
 
The occurrence of these global oscillations depends on the diffusion of
cells, antibodies and virus throughout the whole body array.
Increasing the diffusion constant causes the oscillations to become
more regular, fluctuations in time delay between high virus
concentration phases are smaller, i.e., frequency modulation
decreases.  For low values of the diffusion constant the system is less
synchronized and at a certain threshold value global oscillations
disappear.
 
At the beginning of the cycle virus quickly infects cells and
starts growing.  At this time if the B cells can produce enough
antibodies, before being killed by virus, some {\it islands\/} free of
virus grow randomly.  (A similar phenomenology has been observed in a
very different framework in studying a cellular automaton model for
catalytic oxidation of $CO$ on a $Pt$(100) surface \cite{kapral}.)
The regions free of virus become larger, the islands merge together
and, after some time, they cover the whole lattice.  This is the low
infection phase characterized by low dose of virus and by the
presence of large numbers of antibody over the whole lattice.
Antibodies have a natural time decay, if some virus survives the high
antibody concentration, for example if they find a host cell to infect
before antibodies can eliminate them, they can trigger the infection
process again after most of the antibodies are removed by their natural
decay.
 
We carried out simulations to verify this mechanism by varying the
antibody decay time.  We found that for low values of the antibody
lifetime the high-dose antibody phase is much shorter.  For a lifetime
of one time step, oscillations disappear and we observe a steady state
presence of virus.
 
\subsection{The CTL response\label{ctl}}
 
The next step for implementation of the cellular response in the model
is the introduction of another population of cells: {\it cytotoxic T
lymphocytes\/} (CTL).  In our model CTL are not specific, they have a
fixed probability $P_K$ to interact, and to kill, infected cells.  In
nature the CTL response is specific, however we can think of our
single CTL as equivalent to working with just one kind of antigen,
i.e., the CTLs we introduce are precisely those which interact
with that antigen.
 
In reality CTLs interact with infected cells by recognizing antigenic
peptides presented by class I MHC molecules.  The result is the
destruction of the infected cell and the clonal expansion of the CTLs.
At the same time B cells, helped by lymphocytes, produce antibodies that
can kill or inactivate free virus.  Thus, the cellular and humoral
responses act in parallel against different viral targets.  It is not
known whether the two responses are synchronized but they do have a
similar degree of complexity in their activation and effector steps.
In our simulation the CTL process is simplified and short, since it
consists of a single step.
 
We have performed a series of simulations to evaluate the effect of
CTLs by varying their killing capability $P_K$.  Furthermore, we
assess how the antibody and CTL responses cooperate.  Analyzing
results of the simulations we observe that, of course, the presence of
CTLs helps the immune response.  In some cases CTLs eliminate
infection without the help of antibodies (or use antibodies only in
the last part of the response), and sometimes cooperation of
antibodies and CTL is needed.
 
We have tried to perturb the {\bf O} state of the system without CTL's
by introducing a small number of CTL, as expected we observe that
oscillations are still present for low values of CTL, while they are
eliminated for large amounts of CTL.
 
We ran another series of simulations introducing a
cross-regulation mechanism between CTL and B-cell activation.  In the
real system these processes are mediated by two different sets of
cytokines involving various {\it interleukins\/} (IL) and {\it gamma
interferon }(IFN-$\gamma $).  We have modelled this mechanism by
introducing two different model cytokines called IL and IFN.  The
former is produced by B-T cell binding and down-regulates CTL
division, while the latter accompanies CTL activation and
down-regulates B-cell division (in the human body these processes
exist, but are much more complex and involve many different proteins).
In the presence of this cross-regulation mechanism, the system quickly
chooses one type of response, cooperation is weaker and in only a few
simulations have we observed that both responses are activated at
different times.
 
We have also found an oscillatory state here, but oscillations fade to
an intermediate state between {\bf V} and {\bf IS}; this is a low
infection state where virus is always present in small doses as in a
chronic disease.
 
We also did simulations showing that we can drive the choice of the
system by putting an initial amount of IL or IFN in the lattice at
$t=0$; then one path is enhanced at the beginning which inhibits the
other.
 
\subsection{Comparison to biological data}
 
Upon introducing an infection mechanism, even without CTLs, we obtain
two fixed points: one characterized by indefinite virus growth ({\bf V}),
the other by elimination of the virus ({\bf IS}); these states can be
identified with {\it disease\/} and with {\it recovery\/} from a viral
infection.
 
In the case of evolution to {\bf V}, our simulations show a condition of
{\it immunodeficiency\/}, i.e., a sudden reduction of the
number of cells of the immune system.  We haven't found cases of this
sort of behavior for B cells, but there are a lot of data for a T
helper cells in AIDS ({\it acquired immunodeficiency syndrome\/}).
The mechanism is different because HIV (the virus of AIDS) infects and
kills T cells.
 
In our model we also have an oscillating mode {\bf O}.  Oscillatory
immune responses are found in various experiments {\it in vivo}
and {\it in vitro} \cite{Britton68,delisi1,lawrence}.  Most of these are
devoted to study immune response to a non-proleferating antigen,
such as LPS (lipopolysaccharide, a sugar molecule) \cite{Britton68,delisi1} or
bacterial levan \cite{delisi2}. Cyclic response of T and B cells are found.
The feedback mechanisms involved in the generation of a cyclic response are
generally unknown, however three different feedback mechanism are proposed
to explain these oscillatory patterns. The first one is an antibody feedback
mechanism: after the first antigen injection, B cells produce specific antibodies that bind the antigens and form
antibody-antigen complexes which block further antigenic stimulation
and antibody production.  If the antibody concentration becomes too
low the complexes may dissociate before being eliminated by catabolism
so that the antigens become free and a new cycle can start.  A time
delay differential equation model of this process was proposed by
Grossman {\it et al.} \cite {Grossman80}.
Other feedback mechanisms that have been proposed, are the presence of
auto-anti-idiotypic antibodies according to Jerne's network hypothesis
and the regulatory effects of T helper
cells \cite{rombal,delisi2}.
 
In our system the feedback mechanism is strictly connected to viral
growth, so it is quite different from those previously mentioned,
although the need for regulatory T cells for generating the
oscillations and the presence of high and low antibody concentration
phases are similar in both cases.  In fact in such cases, antigens
(i.e., LPS, bacterial levan) cannot proliferate after the first
injection, while in our case we deal with an infection due to a
microorganism (virus or parasite) that can grow.
 
Closer to our case are oscillations found in some infectious diseases.
Lo, Wear, {\it et al.} \cite{lo} have studied an infection due to {\it
Mycoplasma fermentans} isolated from Kaposi's sarcoma of a patient
with AIDS.  They inoculated silver leaf monkeys with this antigen
and found that all infected animals exhibit oscillations in antigen
concentration in the blood.  Another kind of infectious disease where
an oscillatory pattern can be found is malaria.  Malaria is mainly due
to two parasites {\it Plasmodium vivax} and {\it Plasmodium
falciparum} and it presents periodic sharp episodes of high fever
({\it paroxysms}).  The cyclic behavior of the disease can be
monitored by measuring fever or the concentration of cytokines
involved in the response to the malarial infection \cite{mal}.
 
In presence of CTL and of a simple model of regulation our system can
choose between two different paths: humoral or cell-mediated response.
{\it In vivo \/}this choice is apparently due to the presence of
different patterns of cytokines (see \cite{romagnani}) which arise
from two subclasses of T-helper cells called Th1 and Th2.
Although they are all T-helper cells there is a subtle differentiation
between those that help humoral (Th2) and cellular (Th1)
responses.
 
A model for Th1-Th2 cross-regulation has been proposed by
Fishman and Perelson \cite{Fishman94} based on a system of ordinary
differential equations.  The regulatory mechanism between Th1 and
Th2 is implemented by introducing two species of cytokines.  The
model shows that the immune response is due mainly to one subset of T
helper cells, {\it i.e\/},  they haven't found any stable fixed points
characterized by the contemporaneous (valuable) presence of both kinds
of T helper cells.  The relative efficiency of activation of the
responding Th1 and Th2 cells is the crucial parameter that
drives the choice of the system for one kind of response.  We have
also obtained a similar result in our model by changing the capability
of CTL and B cells to be inhibited by the presence of cytokines.  We
don't have Th1 and Th2 cells in our model, but we have introduced these
two sets of cytokines with regulatory function (called IFN and IL in the model)
as a very preliminary  step towards Th1-Th2 system.  Therefore, as we
have  seen in Section \ref{ctl}, we obtain similar cross-regulation.
 
Fishman and Perelson also studied the evolution of the system by
varying the initial dose of antigen.  They found that, in a particular
region of parameter space, different doses of antigen can induce a
different kind of response.  We haven't found any such behavior in our
model.
 
Another differential equation model of Th1-Th2 regulation has been
proposed by B. Morel, {\it et al.\/}~\cite{Morel92}.  This model gives
a detailed description of the process of interactions between Th1 and
Th2, involving different kinds of cytokines and phases of maturation
of T cells.  In particular they have shown applicability of this
kind of model to immunological experiments {\it in vitro} and the
possibility of using these experiments for a parametric estimation of
the model.  However, they have not yet used their model to simulate an
antigenic attack, so comparison with our results is not possible.

\section{Coupled map model}
\label{sec:maps}
 
We have built a simple system of coupled maps to model the infection
process.  We have preferred to use a coupled map model with respect to
a system of differential equations to preserve the time discreteness
of the cellular automaton model.  There are, of course, important
differences between discrete and continuous time dynamics; coupled
maps reproduce reasonably well the results obtained in the cellular
automaton model in the considered range of parameter values and are
easier and faster to simulate, although, as we will see in our
case, they do not reproduce the oscillatory state.
 
We consider four different species:
 
\begin{itemize}
\item B cells, a certain clone of B cells which is able to interact
with the injected antigen, we call this population $B$;
 
\item infected B cells, we do not consider the possibility that a B
cell can be infected by more than one virus;
 
\item  free virus, $Ag$;
 
\item active B cells, those B cells who have recognized the antigen.
B cells need a second signal from T cells to be activated, this is
modeled by an activation function $T$.  Infected B cells and active
ones are not available for activation.
\end{itemize}
 
We have developed a mean-field model for the kinetics of the species,
considering only averaged concentrations and neglecting diffusion.
The equations that describe the population dynamics are:
 
\begin{equation}  \label{mad}
\begin{array}[b]{l}
B_{n+1}=B_n+s-\mu B_n-\delta Bi_n+d\sigma Ba_n \\
Bi_{n+1}=Bi_n-(\delta +\mu )Bi_n+Pi_n(B_n-B_i) \\
Ba_{n+1}=Ba_n+Pa_n(B_n-(Bi_n+Ba_n))-\lambda Ba_n \\
Ag_{n+1}=Ag_n+r\delta Bi_n-Pi_nB_n-a(1-\sigma )Ba_n+ \\
-Pa_n(B_n-(Bi_n+Ba_n))
\end{array}
\end{equation}
with:
\[
\begin{array}{l}
Pi_n=1-(1-\alpha )^{Ag_n} \\
Pa_n=1-(1-\beta T_n)^{Ag_n} \\
T_n=\epsilon +(1-\exp (-\lambda Ba_n));
\end{array}
\]
$n$ is the time step index and species concentrations are positive real
numbers.
 
In the first equation of the system (\ref{mad}) $s$ is the source term
due to bone marrow production and $\mu B_n$ is the exponential death
term, $\mu $ is the reciprocal of the average lifetime.  $\delta Bi_n$
is the number of infected B cells that die as a consequence of
infection.  $d\sigma Ba_n$ is the clonal expansion term, i.e.,
the B cells that divide after being activated: $d$ is the number
of B cells produced per time step by an activated B cell ($d=1$ if the
time step is the division time) and $ \sigma $ is the fraction of
active B cells that divide at each time step, thus $1-\sigma $ is
the fraction of activated B cells that become plasma cells and secrete
antibodies.
 
The second equation describes the time evolution of the number of
viruses inside a cell $Bi$.  $(\delta +\mu )Bi_n$ is the death term
and $Pi_n(B_n-B_i)$ the infection term.  The probability $Pi_n$ that a cell
is infected by at least one virus is computed as in the cellular
automaton simulation.  We have assumed that the mean free time between
two  cellular and molecular
collisions is small compared to the map time step which is thought to
represent the division time of B cells.  Infection processes are
assumed independent, so let $\alpha $ be the infection probability,
$1-\alpha $ is the probability of not being infected by a given
antigen, $(1-\alpha )^{Ag_n}$ is the probability of not being infected
by any antigen, and $1-(1-\alpha )^{Ag_n}$ is the probability of being
infected by at least one antigen per B cell and $ (1-(1-\alpha
)^{Ag_n})(B_n-B_i)$ is the total number of infected B cells per time step,
because only uninfected B cells are available for infection.
 
The third equation describes the evolution of active B cells.  $
Pa_n(B_n-(Bi_n+Ba_n))$ is the growth term due to B cell activation;
the probability for a B cell to be activated by at least one antigen
($Pa_n$) is computed in the same manner as $Pi_n$, but the probability
of activation is $ \beta $ times a certain activation function ($T_n$)
that represents the effect of T cells.  $T_n$ introduces a positive
feedback.  At the beginning of the response the number of active T
cells is small; then, because of B cell-T cell binding followed by
B-cell activation and T-cell duplication, the effect becomes larger.  To
represent this we have chosen the function:
 
\[
T_n=\epsilon +(1-\exp (-\lambda Ba_n));
\]
with $\epsilon $ the minimal activation probability in absence of
active B cells.  We consider only those B cells that are not infected
or yet activated to be available for activation.  The last term of the
equation, $-\lambda Ba_n$, is the inactivation term, where $\lambda $
is the reciprocal of the average activation lifetime.  We neglect the
possibility that active B cells die ($ \lambda \gg \mu $).
 
The last equation of the system (\ref{mad}) is the evolution equation
for free antigens, the source term is due to death of infected B
cells, each of them releasing $r$ antigens.  Some antigens enter the
cells because of the infection process ($-Pi_n(B_n-B_i)$) and of the
internalization following B-cell activation
($-Pa_n(B_n-(Bi_n+Ba_n))$).  Each plasma B cell produces $a$
antibodies, that eliminate $a(1-\sigma )Ba_n$ antigens.
 
We have to impose some more constraints on (\ref{mad}), because, for
example, $Pi_n(B_n-B_i),$ the number of B cells that are infected, has
to be smaller than both $B_n-B_i$ (and this is always verified) and
$Ag_n$, this last condition is implemented by the $min$ function.
The same argument is true for other terms.  After having imposed these
constraints we obtain the system:
 
\begin{equation}
\begin{array}{l}
B_{n+1}=B_n+s-\mu B_n-\min (\delta Bi_n,B_n)+d\sigma Ba_n \\
Bi_{n+1}=Bi_n-\delta Bi_n+\min (Pi_nB_n-B_i,Ag_n) \\
Ba_{n+1}=Ba_{n+1}+\min (Pa_n(B_n-(Bi_n+Ba_n)),Ag_n)-\lambda Ba_n \\
Ag_{n+1}=Ag_n+r\delta Bi_n-\min (Pi_nB_n,Ag_n)-a(1-\sigma )Ba_n+ \\
-\max (\min (Pa_n(B_n-(Bi_n+Ba_n)),Ag_n),0)
\end{array}
\label{com}
\end{equation}
 
\subsection{Numerical results}
 
We have studied the response to an antigenic stimulation.  Thus our
typical initial condition in the ($B$, $Bi$, $Ba$, $Ag$) space is
($B_o$, $0$, $0$, $Ag_o$).  We carried out simulations varying $r$
and $\alpha$, the parameters corresponding to $P_I$ and $V_{\max }$
in the cellular automaton model.  Extensive simulations show that in
all these cases the system evolves to one of these two fixed points:
 
\begin{itemize}
\item[-]  ($\frac s\mu $, $0$, $0$, $0$) with antigen elimination which
we call the immune state ($IS)$;
 
\item[-] ($\frac s{\mu +1}$, $\frac s{\left( \delta +\mu \right)
\left( \mu +1\right) }$, $0$, $+\infty $) with antigen growth which we
call the diseased state ($V$).
\end{itemize}
 
$IS$ is reached for small values of $r$ and $\alpha $, otherwise the
system evolves to $V$.  We have verified numerically that the system
(\ref{com}) has other fixed points but they are all locally unstable,
also $IS$ is locally unstable in some range of values of $\alpha $ and
$r$. $IS$ lies on the border of the domain of definitions of our dynamic
variables (i.e. positive real numbers), so we have studied its stabilty
considering only allowed (i.e. positive) perturbations.

The mechanism by which $IS$ is reached is inherently linked to the
time discreteness of the dynamics and to the fact that the stable
manifold of $IS$, $(B,0,0,0)$, belongs to the boundary of the domain.
The rule by which we update the values of concentrations when they
become negative is to set them to zero, which is a boundary value, and
it may then be attracted to $IS$ if it happens to be close to the
stable manifold.  This kind of mechanism represents closely what
happens in the cellular automaton model, where the discreteness of the
state variable may lead to the disappearance of a species in one time
step by fluctuations.

We have fixed a value for $\alpha $ ($\alpha =0.05$) and studied the
modifications of the basin of attraction of $IS$ with initial
conditions of the class ($B_o$, $0$, $0$, $Ag_o$) ({\it antigenic
stimulation\/}), varying $r$.
 
We have observed that $IS$ is inside its basin for low values of $r$,
so $IS$ is stable for ($\frac s\mu +\epsilon $, $0$, $0$, $\epsilon
^{\prime }$) like perturbations, with $\epsilon $, $\epsilon ^{\prime
}\ll \frac s\mu $. Vice versa it is at the border for $r>r_o$
($r_o\simeq 11.045$ with $ \alpha =0.05$) and ($\frac
s\mu+\epsilon$, $0$, $0$, $\epsilon ^{\prime }$) belongs to the basin of
attraction of $V$.  In Fig.~\ref{attr} we plot the basins of
attraction of $IS$ and $V$ in the ($B$, $0$, $0$, $Ag$) plane under {\it
antigenic stimulation\/}; the border between the two basins moves with
$r$.  We observe that the border curve in the ($B$,$Ag$) plane is not
monotonic with B. Analysing the temporal behaviour of the four
populations for various initial $B$ values, with other parameters
fixed, we find the ratio $\frac{Bi}{Ba}$ to be the crucial parameter
for the choice between evolution to one fixed point or the other.  In
fact, in the $IS$ case, after an initial transient, $\frac{Bi}{Ba}$ is
below 1, while in the $V$ case it is greater than one.  Both
populations, $Ba$ and $Bi$, increase with the number of initial B
cells, but their ratio $\frac{Bi}{Ba}$ after the initial transient is
not monotonic in the initial value of B. This causes the dip in the
curve of Fig.~\ref{attr}.
 
The loss of stability of $IS$ is due to the fact that for $r<r_o$ an
orbit started close to $IS$ returns near the stable manifold of $IS$,
while for $  r>r_o$ the orbit goes away to $V$.  In fact, if we plot
the projections of some orbits, starting near $IS$, on the ($B$,$Ag$)
plane we can see (Fig.~\ref{orb})
that close to the fixed point orbits with
different values of $r$ are similar but then, for $r>r_o$, the orbits
evolve to $V$, while for $r<r_o$ we have a sort of homoclinic
phenomenon because the orbit reaches the stable manifold of $IS$ (the
$B$ axis) and so $IS$ itself.  The $\omega $-limit set (see Ref.
\cite{Guck} pag.235) of
an $IS$ neighborhood is $IS$ itself for $r<r_o$, $IS\cup V$ for
$r=r_o$ and $V$ in the case of $r>r_o$ (except the $B$ axis always
belongs to the stable manifold of $IS$).  We are presumably in the
presence of a {\it global bifurcation \/}since the transition
$IS\Rightarrow V$ is not due to the loss of local stability but to
the behavior of the orbits far from the fixed point.
 
It is also interesting to observe that the dynamics of the relevant
species show a similar behavior to that found in the cellular
automaton simulation.  In Fig.~\ref{conc} we show the time evolution
of B cells and Ag concentrations in both the $IS$ and the $V$ case.
 
Therefore, using a simple mean field model we have found some features of
the cellular automaton simulation, i.e.:
 
\begin{itemize}
\item[-]  evolution to $IS$ or $V$;
 
\item[-] same behavior when varying the parameters corresponding to
$P_i$ and $ V_{\max }$;
 
\item[-]  similar behavior of populations in the $IS$ and $V$ cases.
\end{itemize}
 
We have not found oscillatory states (cycles or limit cycles) at the
border between $IS$ and $V$.  Since in Section \ref{infect} diffusion
is shown to play an essential role in establishing oscillations,
we presume that spatial effects are relevant and could be added to the
mean field model in a sort of diffusively coupled map lattice.

\section{Conclusions}
 
We have extended the Celada-Seiden (CS) model to include cellular
response and we have studied the dynamics of an infectious disease and
the cross-regulation between humoral and cellular response.  With
respect to the CS model we have included cytotoxic T lymphocytes
(killer cells) and some generalized cytokines; moreover, the infecting
virus is recognized by the immune system as antigen and is attacked
both by antibodies outside an infected cell and by killer cells inside
the infected cell.  To our knowledge this is a first attempt to
simultaneously treat the two main branches of specific immune
response.  Although a number of simplifications were made (e.g.,
non-specificity of T killer cells), some interesting results were
obtained.
 
We performed a number of simulations and compared our results with
{\it in vivo\/} and {\it in vitro\/} experiments.  By varying the
probability of viral infection and the number of viral particles
released after cell lysis we obtained two fixed points towards which
the systems evolve.  These are identified with disease or recovery
after the immune response.  At the border between these two states we
find persistent synchronized oscillations of all populations.  This
result is so robust as to be also present in the absence of killer
cells.  The introduction of killer cells obviously improves the immune
response and, through cross-regulation mechanisms, biases the response
towards one of the two branches, as is observed {\it in vivo}.  We
have also found that fluctuations can destroy the oscillating mode and
drive the system to one of the stable fixed points.
 
We also discussed a mean field model based on coupled maps to study an
infection due to a virus-like agent, with the aim of understanding the
main feature of the CS simulations, i.e.,
evolution towards disease or
recovery.  This very simple model finds the same fixed points, and the
temporal behavior of the concentrations of the pertinent entities are
also in agreement with the CS model.  However, the dynamical behavior
connected with spatial extension is lost, as is the oscillating mode.
The mean field approximation introduced with coupled maps destroys the
oscillations.  This shows that the spatial effects introduced by the
cellular automaton model are crucial to obtain the oscillatory state,
which is quite an important and interesting new feature.  The mean
field limit behavior of the coupled map model can be reached by
increasing the diffusion constant of CA model.
 
Our model is just a first attempt towards a complete implementation of
cellular response in which both responses are simulated in specific
manner, as in the real system.  However the results presented in this
paper show the wide possibilities of a microscopic approach to the
study of immune response.
 
\section{Acknowledgments}
 
We acknowledge the Institute for Scientific Interchange in Torino for
hospitality, the Department of Biology and Animal Genetics
(Universit\`a di Firenze), IBM and the Hospital for Joint Diseases
(New York, USA) for finacial support (M.B.); R.Kapral and F. Bagnoli
for discussions.
 

\newpage
 
\begin{figure}
\caption{a) Entities in the Celada-Seiden model simulation for the
case of $8$ bits.  b) Process of recognition, internalization and
presentation of antigen.  Numbers indicate decimal values of the
binary strings.}
\label{cells}
\end{figure}
 
\begin{figure}
\caption{A typical immunization  experiment.}
\label{immsim}
\end{figure}
 
\begin{figure}
\caption{Evolution of the population of B cells and antigens for
three cases of $V_{\max }$ with $P_i=0.05$.}
\label{evol}
\end{figure}
 
\begin{figure}
\caption{ The separatrix of the basins of attraction of $IS$ and $V$
for three different values of the number of viruses injected back into
the system by virus induced cell death ($r$).  The lower right is the
basin of $IS$, the upper left is the basin of $V$.}
\label{attr}
\end{figure}
 
\begin{figure}
\caption{
Projections on ($B$,$Ag$) plane of some orbits for
different values of $r$ and same initial condition ($\frac s\mu $,$\,0$,$
10^{-3}$,$0$).}
\label{orb}
\end{figure}
 
\begin{figure}
\caption{Evolution of concentration of B cells and antigens for the
cases of $IS$ and $V$.}
\label{conc}
\end{figure}
 
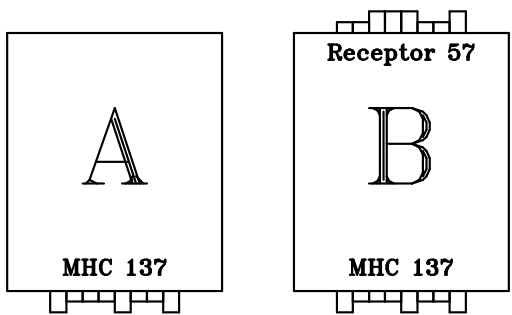
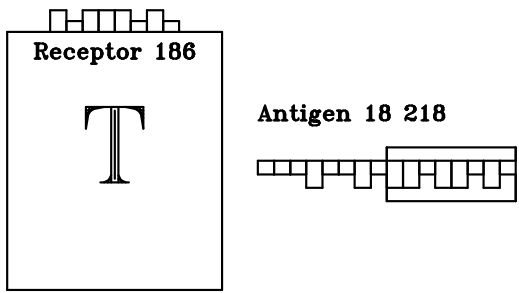
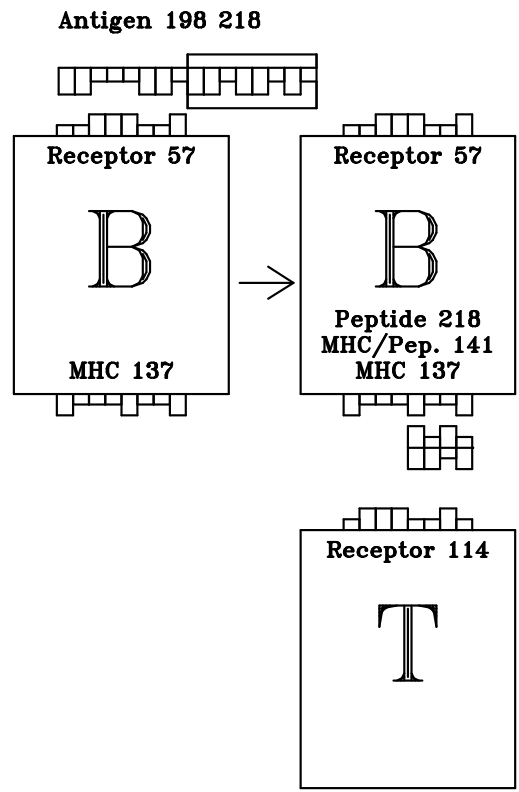
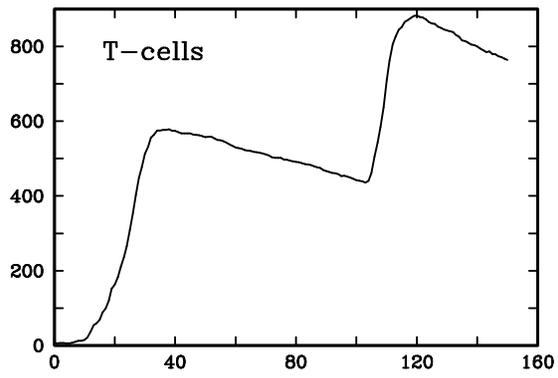
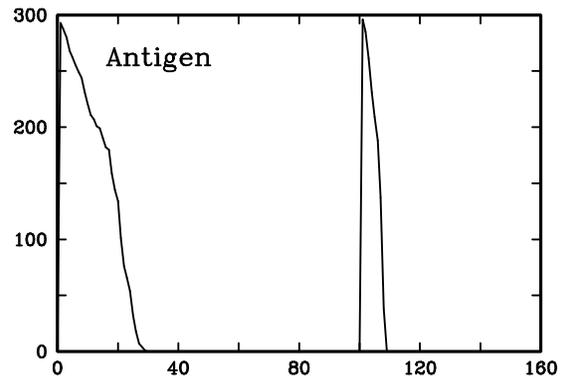
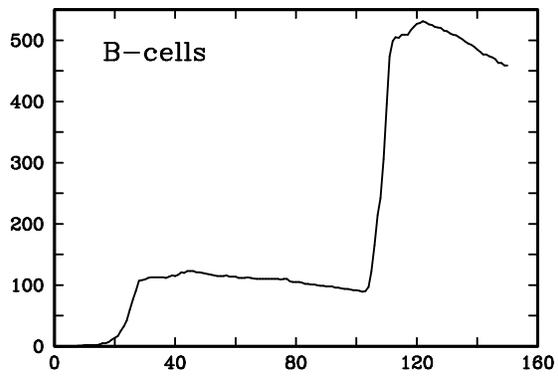
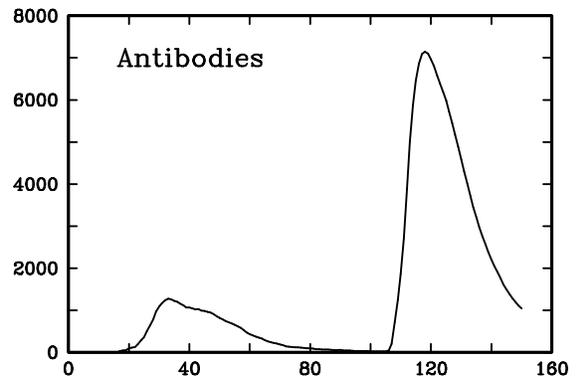
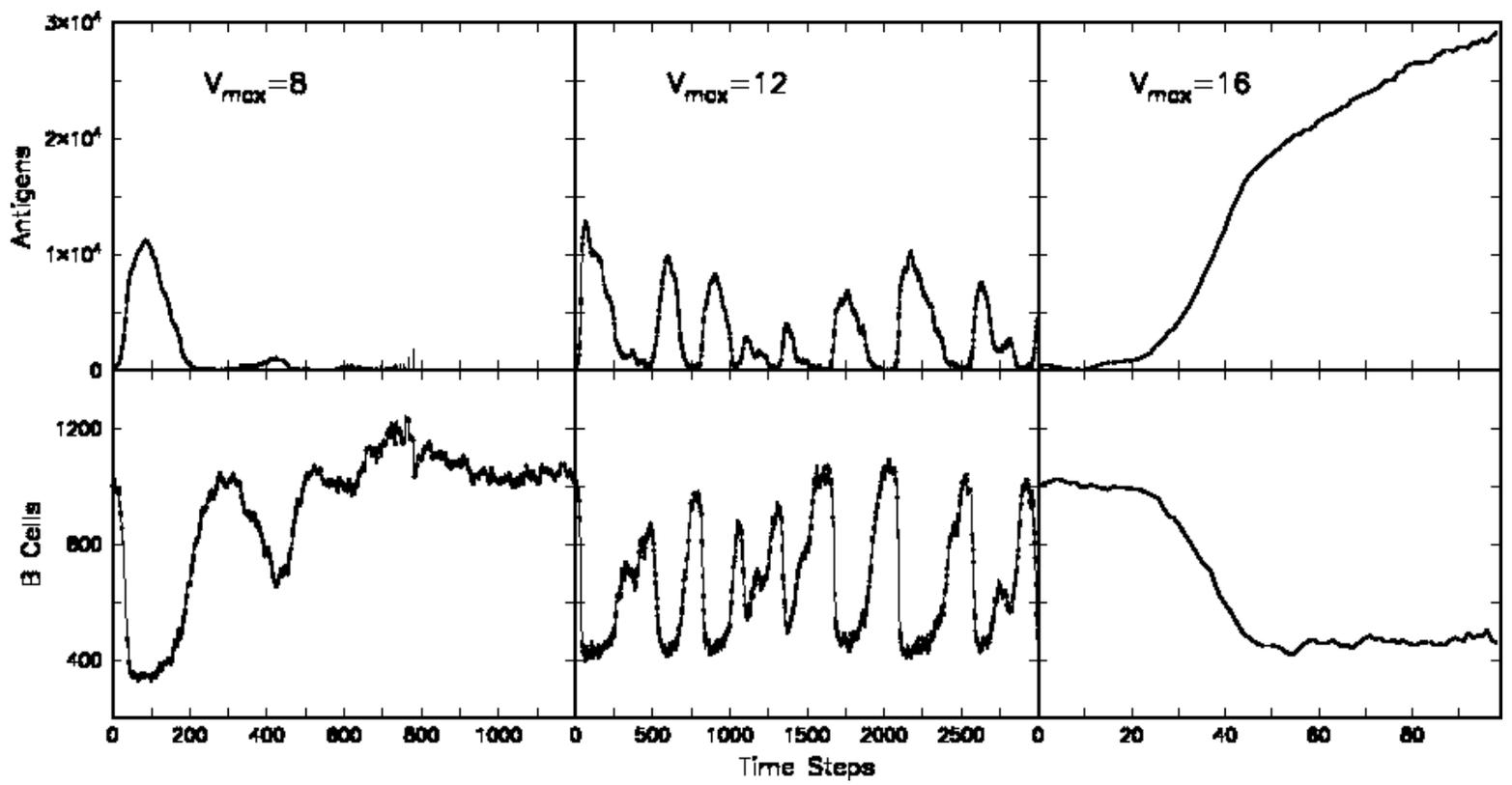
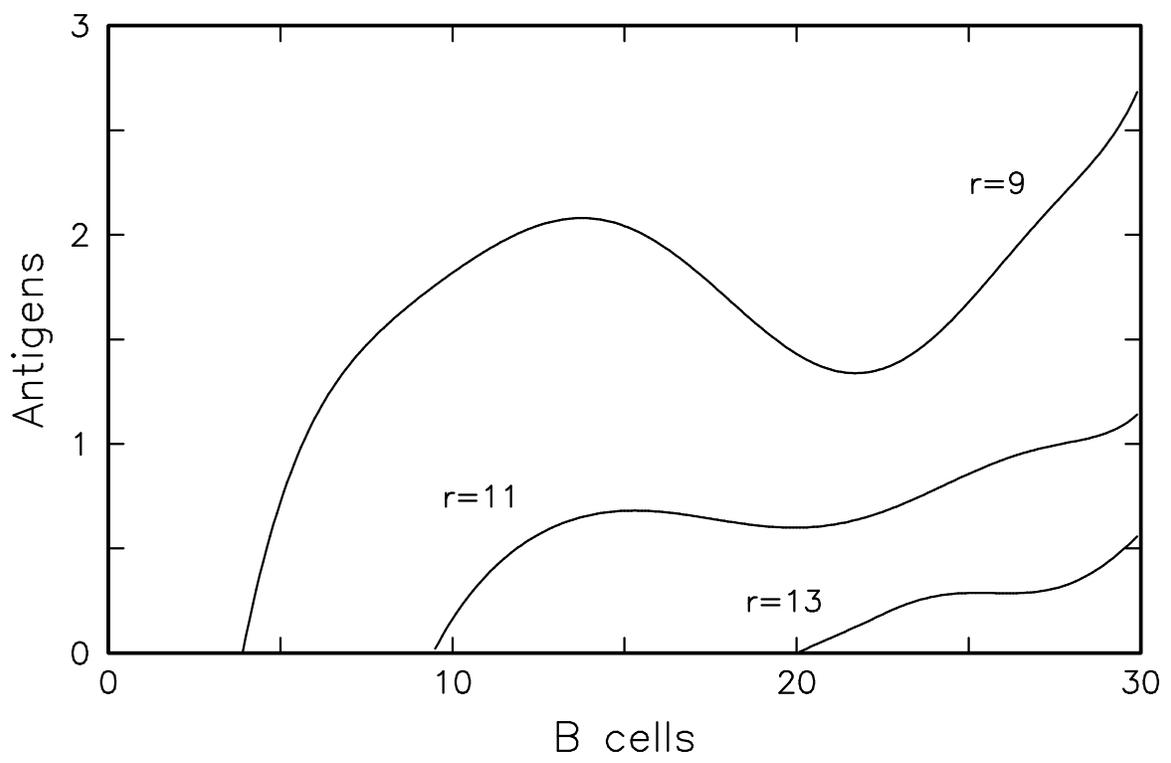
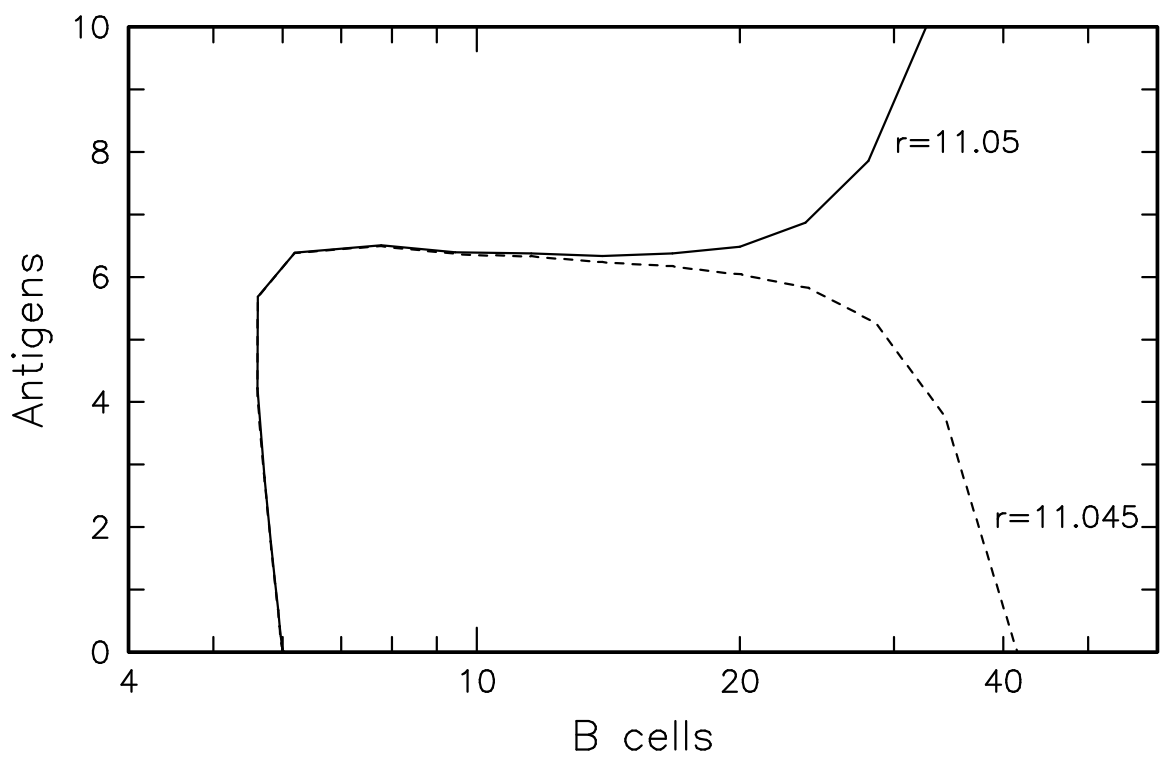
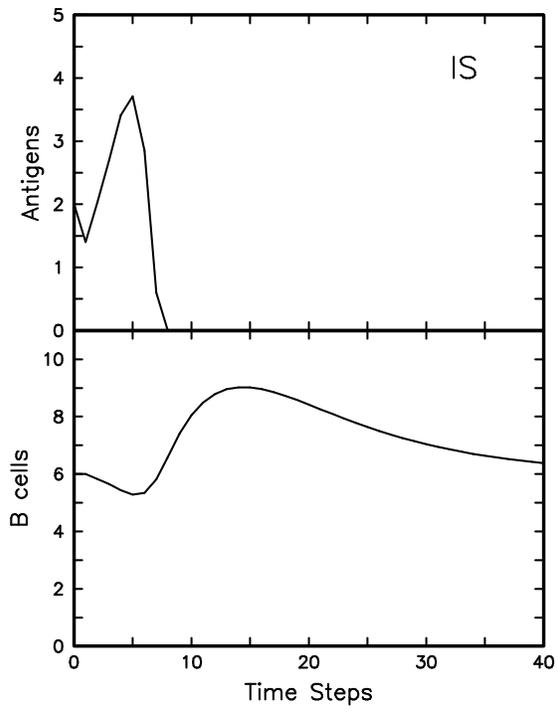
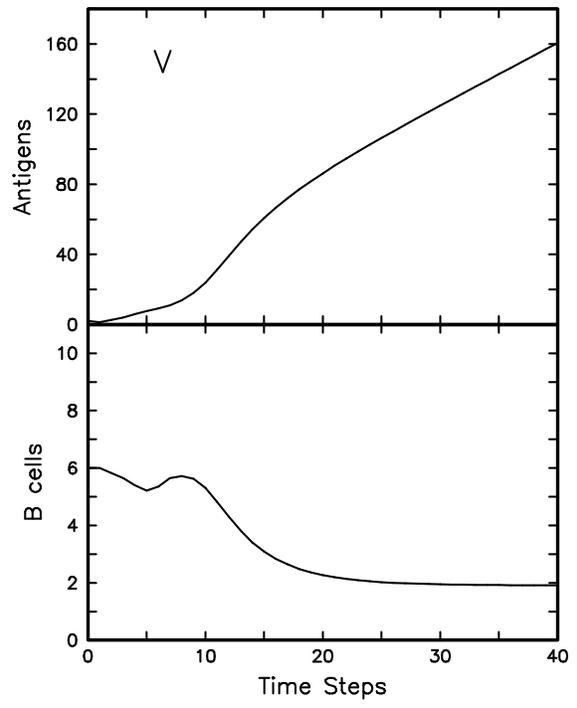
\end{document}